\documentclass[prd,twocolumn,showpacs,preprintnumbers,amsmath,amssymb,superscriptaddress,floatfix,nofootinbib]{revtex4}
\usepackage{amssymb}
\usepackage{amsmath}
\usepackage{amsfonts}
\usepackage{epsfig}
\usepackage{graphicx}
\usepackage{tabularx}
\usepackage{latexsym}
\usepackage{subfigure}
\usepackage{verbatim}
\usepackage{color}
\usepackage{braket}
\usepackage{multirow}

\usepackage[normalem]{ulem}

\renewcommand\sout{\bgroup \color{red} \ULdepth=-.5ex \ULset}

\begin{document}

\title{Six-quark structure of $d^*(2380)$ in chiral constituent quark model}

\author{Qi-Fang L\"{u}} \email{lvqifang@ihep.ac.cn}
\affiliation{Institute of High Energy Physics, Chinese Academy of Sciences, Beijing 100049, China}
\affiliation{School of Physical Sciences, University of Chinese
Academy of Sciences, Beijing 101408, China}
\author{Fei Huang}
\affiliation{School of Physical Sciences, University of Chinese
Academy of Sciences, Beijing 101408, China}
\author{Yu-Bing Dong}
\affiliation{Institute of High Energy Physics, Chinese Academy of
Sciences, Beijing 100049, China}
\affiliation{School of Physical Sciences, University of Chinese
Academy of Sciences, Beijing 101408, China}
\affiliation{Theoretical Physics
Center for Science Facilities (TPCSF), CAS, Beijing 100049, China}
\author{Peng-Nian Shen}
\affiliation{College of Physics and Technology, Guangxi Normal
University, Guilin 541004, China}
\affiliation{Institute of High
Energy Physics, Chinese Academy of Sciences, Beijing 100049, China}
\affiliation{Theoretical Physics Center for Science Facilities
(TPCSF), CAS, Beijing 100049, China}
\author{Zong-Ye Zhang}
\affiliation{Institute of High Energy
Physics, Chinese Academy of Sciences, Beijing 100049, China}
\affiliation{Theoretical Physics Center for Science Facilities
(TPCSF), CAS, Beijing 100049, China}

\begin{abstract}

The structure of $d^*(2380)$ is re-studied with the single cluster
structure in the chiral SU(3) quark model which has successfully
been employed to explain the $N-N$ scattering data and the binding
energy of deuteron. The binding behavior of such a six quark system
is solved by using a variational method. The trial wave function is
chosen to be a combination of a basic spherical symmetric component
of $[(0s)^6]_{orb}$ in the orbital space with $0\hbar\omega$
excitation and an inner structural deformation component of
$[(0s)^5(1s)]_{orb}$ and $[(0s)^4(0p)^2]_{orb}$ in the orbital space
with $2\hbar\omega$ excitation, both of which are in the spatial [6]
symmetry. It is shown that the mass of the system is about $2356$
MeV, which is qualitative consistent with the result both from the
two-cluster configuration calculation and from the data measured by
the WASA Collaborations. This result tells us that as long as the
medium-range interaction due to the chiral symmetry consideration is
properly introduced, the mass of system will be reduced in a rather
large extent. It also implies that the observed $d^*$ is a six-quark
bound state with respect to the $\Delta\Delta$ threshold, which
again supports the conclusion that $d^*$ is a hexaquark dominant
state.
\end{abstract}
\maketitle

\section{Introduction}{\label{introduction}}

During past years, a resonant structure $d^*(2380)$ has been
reported in the double-pion fusion reactions $pn \to d \pi^0 \pi^0$
and $pn \to d \pi^+ \pi^-$ by the WASA-at-COSY collaborations
\cite{Bashkanov:2008ih,Adlarson:2011bh}. Later on, this resonance
has also been observed in $pn \to pn \pi^0 \pi^0$, $pn \to pp \pi^-
\pi^0$, $pd \to\,^3He\,\,\pi^0 \pi^0$, $pd \to\,^3He\,\,\pi^+
\pi^-$, $dd \to\,^4He\,\,\pi^0 \pi^0$, and $dd \to\,^4He\,\,\pi^+
\pi^-$
reactions\cite{Keleta:2009wb,Adlarson:2012au,Adlarson:2013usl,
Adlarson:2014tcn,Adlarson:2014pxj,Adlarson:2014xmp}, and further
confirmed by incorporating the newly measured analyzing power data
into the partial wave
analysis\cite{Adlarson:2014pxj,Adlarson:2014ozl}. The data show that
$d^*(2380)$ has a mass of 2380 MeV, a width of $\Gamma \approx 70
\rm{MeV}$, and an isospin-spin-parity of
$I(J^P)=0(3^+)$~\cite{Clement:2016vnl}.

Since the mass of $d^*(2380)$ is away from the thresholds of the
$\Delta \Delta$, $\Delta N \pi$, $NN \pi \pi$ channels, the
threshold effect is expected to be smaller than that in some exotic
XYZ states\cite{Zhusl:2016,Olsen:2015zcy}. The structural
uncertainty in studying $d^*(2380)$ would be much smaller. On the
other hand, although observed mass is higher but not much higher
than the $\Delta N \pi$ and $NN \pi \pi$ thresholds, its width is
only 70 MeV, which is much smaller than the width of two $\Delta$s.
The fact that the width of $d^*(2380)$ is remarkably small excludes
the scenario of the na\"ive $\Delta \Delta$ molecular structure where
the $\Delta$s are color singlet particles and indicates that the
effect of the hidden-color channel should be
significant\cite{Bashkanov:2013cla,Yuan:1999pg}. Due to these
extraordinary properties of $d^*(2380)$, it becomes a good platform
to reveal some information about the new structure in the hadronic
system.

The properties of dibaryon states were firstly discussed by Dyson
and Xuong in 1964 in the framework of SU(6) symmetry where no
dynamics is considered \cite{Dyson:1964xwa}. Since then, various
theoretical investigations on dibaryon have been performed.
Recently, Gal and Garcilazo studied the $\pi N \Delta$ system in a
Faddeev type three-body calculation and dynamically generated a pole
where its mass and width are close to the data of WASA, although
some approximations were employed \cite{Gal:2013dca,Gal:2014zia}. In
Ref.~\cite{Huang:2013nba}, H. Huang, et al, investigated the binding
behavior of the $\Delta\Delta$ system in a coupled channel
calculation in the framework of the chiral SU(2) model and obtained
a binding energy of about $71$MeV and a width of about $150$MeV,
which is much larger than the reported data. Even in a QCD sum rule
calculation, one can also get a mass of $2.4\pm0.2$
GeV\cite{Chen:2014vha}. However, a recent calculation by using a
constitute quark model with the one-gluon-exchange (OGE) and
confinement interactions only showed that the six $u$-$d$ quark
system with the [6] symmetry in the orbital space should not be
bound~\cite{Park:2015nha}.

It is noteworthy that in a much earlier calculation in the chiral
SU(3) quark model, the binding property of the $\Delta\Delta$ system
with $I(J^P)=I(S^P)=0{(3^+)}$, where $I, J$(or $S$), and $P$ stand
for the isospin, spin, and parity, respectively, was studied by
including a hidden color (CC) component , and a bound state with a
binding energy of $40-80 \rm{MeV}$ relative to the threshold of the
$\Delta\Delta$ channel was predicted\cite{Yuan:1999pg,Dai:2005kt}.
After the new discovery by the WASA-at-COSY collaborations, more
detailed calculations for such a state have been performed on the
base of the chiral SU(3) quark model and extended chiral SU(3) quark
model\cite{Huang:2014kja,Dong:2015cxa,Huang:2015nja}. In the
framework of the Resonating Group Method (RGM), the mass and wave
function of the state are obtained by dynamically solving the
coupled-channel equations where the coupling of the $\Delta\Delta$
channel with a hidden color channel has been considered. The partial
decay widths of the $d^* \to d~ \pi \pi$, $d^* \to NN~ \pi \pi$ and
$d^* \to NN~ \pi $ processes are evaluated in terms of the extracted
wave function, and the total width of about $71$MeV of $d^*$, which
coincides with the averaged experimental value of $75$MeV, is
obtained~\cite{Dong:2015cxa,Dong1,Dong2}. It is shown that due to a
large CC component in the system, the resultant mass and width are
compatible with the data, namely, such a component plays an
essential role in interpreting the observed characters of $d^*$,
especially its narrow width. Thus, one would conjecture that
$d^*(2380)$ might be a hexaquark dominated exotic state.

Inspired by a large CC component in $d^*(2380)$ and a small size in
the coordinate space, it is reasonable to study the $(IS)=(03)$ six
quark system in an alternative model space with a single cluster
configuration (SCC). On the other hand, according to Harvey's
relation from the group theory~\cite{Harvey:1981}, in a six quark
system, single-cluster configurations and two-cluster configurations
(TCC) can transform each other via Fierz transformation. In this
sense, if $d^*(2380)$ is a hexaquark dominated state, the major
characters obtained in the TCC calculation, say the binding behavior
and the narrow width, should also appear in the SCC calculation.
This is because that by re-arranging the form of the wave function
obtained in the TCC calculation, one finds that the main component
in TCC is a genuine six-quark configuration
$(0s)^6[6]_{orb}[111111]_{SIC}$, or called hexaquark configuration.
Therefore, the main aim of this paper is to see whether SCC has
similar properties as those obtained in the TCC calculation. In this
work, we would re-study such a six quark system in a SCC calculation
with the same chiral SU(3) quark model and the same model
parameters. The trivial wave function consists of a
$(0s)^6[6]_{orb}[111111]_{SIC}$ component together with a component
with a 2$\hbar\omega$-excitation, which is orthogonal to the wave
function of the excited center of mass motion.

It should particularly be emphasized that due to the importance of
the chiral symmetry in the strong interaction, such a symmetry
should be restored in the Lagrangian of the hadronic system, which
leads to the well-known $\sigma$ model~\cite{GL:1960nc}. In the
QCD-inspired constituent quark model, the spontaneous symmetry
breaking of vacuum generates the Goldstone boson, and consequently,
the constituent quark mass. Based on this Goldstone theory, the
Goldstone boson has to be introduced if a constituent quark model is
adopted. That is why a chiral SU(3) constituent quark model was
proposed. With such a model, most data of the ground state
properties of baryons, the baryon spectrum, the baryon-baryon
scattering phase shifts and cross sections, and some binding
behaviors of two-hadron systems, for instance, deuteron and $H$
particle, etc. can be explained in quite good
extent~\cite{Zhang:1997ny}, although this model is somehow a
preliminary attempt to modeling the non-perturbative effect of QCD
(NPQCD). However, if in a constituent quark model, the inter-quark
interaction includes only the OGE and confinement terms (naive OGE
quark model), the scattering and binding behaviors between nucleons
might not be reasonably explained, because the medium-range NPQCD,
which is described by the Goldstone boson exchange in our chiral
constituent quark model, is missing. Therefore, another goal in this
paper is to see that if the interactions arising from the Goldstone
boson exchange are incorporated into the na\"ive OGE quark model,
whether the conclusion in Ref.~\cite{Park:2015nha} could be changed.

The paper is organized as follows. In Sect.\ref{model}, the
formalism for both interaction and wave functions is briefly
introduced. The results and discussions are given in
Sect.\ref{results}. Finally, a short summary is provided in
Sect.\ref{summary}.

\section{Brief formulism}{\label{model}}
\subsection{Interaction}{\label{interaction}}

The interactive Lagrangian between the quark and chiral field in the
chiral SU(3) constituent quark model can be written as
\begin{eqnarray}
{\cal L}_I^{ch}&=&-g_{ch} \bar \psi (\sum_{a=0}^8 \lambda_a \sigma_a +
i \gamma_5 \sum_{a=0}^8 \lambda_a \pi_a) \psi,
\end{eqnarray}
where $g_{ch}$ is the coupling constant of quark with the chiral
field, $\psi$ is the quark field, and $\sigma_a$ and $\pi_a$
$(a=0,1,...,8)$ are the scalar and pseudo-scalar nonet chiral
fields, respectively. Then, the interactive Hamiltonian can be
obtained by,
\begin{eqnarray}
{\cal H}_I^{ch}&=&g_{ch} F(\boldsymbol{q}^2) \bar \psi (\sum_{a=0}^8
\lambda_a \sigma_a +i \gamma_5 \sum_{a=0}^8 \lambda_a \pi_a) \psi,
\end{eqnarray}
where the form factor $F(\boldsymbol{q}^2)$ is introduced to imitate
the structures of the chiral fields. The form of
$F(\boldsymbol{q}^2)$ is usually taken as

\begin{eqnarray}
F(\boldsymbol{q}^2)= (\frac{\Lambda^2}{\Lambda^2+\boldsymbol{q}^2})^{1/2},
\end{eqnarray}
with $\Lambda$ being the cutoff mass, which corresponds to the scale
of the chiral symmetry
breaking\cite{Kusainov:1991vn,Buchmann:1991cy,Henley:1990kw}. From
this Hamiltonian, the chiral field caused quark-quark interaction
$V^{\sigma_a}$ and $V^{\pi_a}$, which mainly provide the
medium-range interaction from NPQCD, can easily be derived. To
reasonably describe the short-range interaction from the
perturbative QCD (pQCD), one-gluon-exchange (OGE) interaction
$V^{OGE}$ is still employed. It should be emphasized that double
counting does not occur between the OGE and chiral field caused
interactions, because the former is a short-range interaction from
pQCD and the latter describes the medium-range interaction from
NPQCD, respectively. Meanwhile, a phenomenological confining
potential $V^{conf}$ is again adopted to account for the long-range
interaction from NPQCD. Consequently, the total Hamiltonian of a
six-quark system in the chiral SU(3) quark model, it can be given by
\begin{eqnarray}
H=\sum_{i=1}^6 T_i -T_G + \sum_{j>i=1}^6(V_{ij}^{OGE}
+V_{ij}^{conf}+V_{ij}^{ch}),
\end{eqnarray}
where the $T_i$ and $T_G$ are the kinetic energy operators of the
$i$-th quark and the center of mass motion (CM), respectively,
$V_{ij}^{\alpha}$ with $\alpha={\rm OGE, conf, ch}$ denote the OGE,
confinement, and chiral field induced interactions between the
$i$-th and $j-$th quarks, respectively,
\begin{eqnarray}
V_{ij}^{ch}=\sum_{a=1}^8 V_{ij}^{\sigma_a} + \sum_{a=1}^8 V_{ij}^{\pi_a}.
\end{eqnarray}
The explicit expressions of these potentials can be found in
Ref.\cite{Zhang:1997ny}. In this work, a chiral SU(3) quark model
with a linear confining potential is employed to reveal the binding
character of the concerned six-quark system with a SCC structure,
and corresponding model parameters are listed in Table~\ref{tab1},
\begin{table}
\begin{center}
\caption{ \label{tab1} Model parameters of chiral SU(3) quark model
with linear confinement. The masses of exchanged mesons are
$m_{\sigma^\prime} = 980 ~ \rm{MeV}$, $m_\epsilon = 980 ~ \rm{MeV}$,
$m_\pi = 138 ~ \rm{MeV}$, $m_\eta = 549 ~ \rm{MeV}$,
$m_{\eta^\prime} = 957 ~ \rm{MeV}$, respectively. The cutoff mass is
$\Lambda = 1100 ~ \rm{MeV}$.} \scriptsize
\begin{tabular}{cccccccc}
\hline\hline
   & & & & &\\
  $b_u$(fm)          & $m_u$(MeV)            & ~~~$g_u^2$~~~       & ~~~~$g_{ch}$~   &
 $m_\sigma$(MeV)   & $a^c_{uu}$(MeV/fm)  &
$a^{c0}_{uu}$(MeV)\\\\ \hline
  & & & & & \\
  0.5              & 313                 & 0.766       & ~~~2.621    & 595
 & 87.5              &  -77.4 \\
  & & & & & \\
\hline\hline
\end{tabular}
\end{center}
\end{table}
in which the coupling constant $g_{ch}$ of the chiral field with
quarks is determined by the experimental value of the $NN\pi$
coupling constant $g_{NN\pi}$, the coupling constant $g_{u}$ of the
gluon with quarks is fixed by the mass difference between $N$ and
$\Delta$, the confining strength $a_{uu}^c$ of the OGE potential is
obtained by satisfying the stability condition of the nucleon ($N$),
the zero-point energy $a_{uu}^{c0}$ is fixed by the mass of $N$, the
masses of the exchanged bosons are chosen from the empirical masses
of relevant mesons and by fitting the data of the $N$-$N$ scattering
and the binding energy of the deuteron. These values are exactly the
same as those in our previous two channel RGM calculations except
the values of $a^c_{uu}$ and $a^{c0}_{uu}$, because here the
quadratic confinement is replaced by a linear
one\cite{Huang:2015nja}. The reason for choosing a linear confining
potential is that due to the NPQCD effect, the confining potential
prefers a linear form rather than a quadratic one. According to the
lattice calculation, it even tends to a color screened form whose
strength is weaker than that of the linear one at the larger
separation between quarks. Moreover, in the hadronic spectrum study,
because the mass scale is about GeV, the NPQCD effect is surely
nonnegligible, as a consequence, the spectrum will be sensitive to
the form of the confining potential, especially in the SCC
calculation. In order to provide a meaningful prediction about the
binding behavior of $d^*$, it is better to take a linear form or
even color screened form for the confining potential. It should be
specially mentioned that in the nucleon-nucleon case, since the
inter-cluster interaction intervenes between two color singlets, the
choice of the different confining potential will not cause visible
effects in the $N-N$ interaction. Namely, using a linear confining
potential instead of a quadratic one will not affect either
scattering phase shifts or binding results between
nucleons\cite{Zhang:1997ny,Huang:2015nja}. Based on the above
reasoning, we can ensure that the model with a linear confining
potential still possesses the prediction power.

\subsection{Wave function}
\begin{figure}[htbp]
\centering
\includegraphics [width=4.2cm, height=3.5cm]{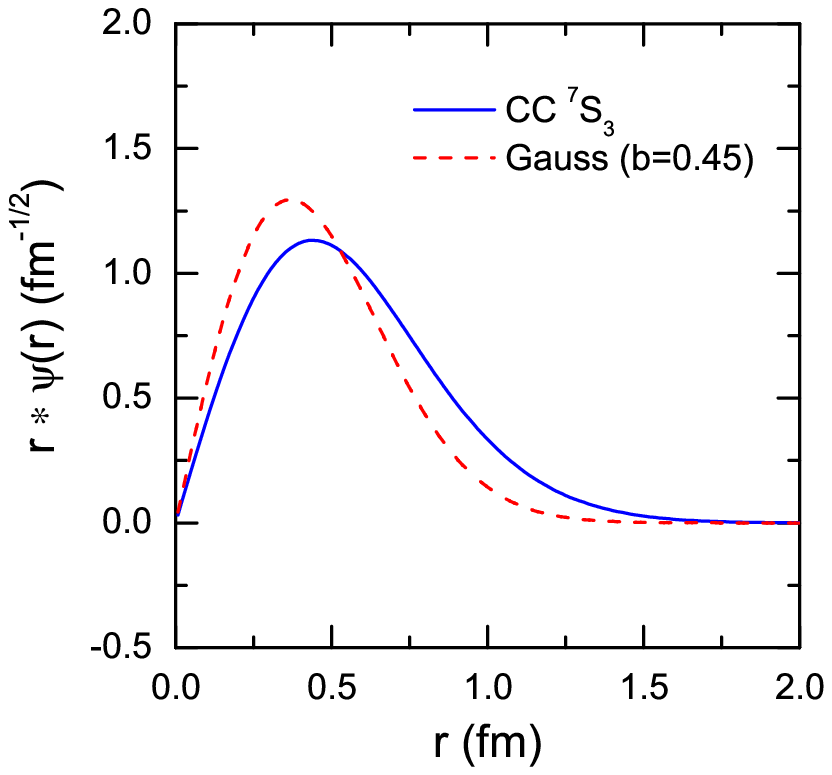}
\includegraphics [width=4.2cm, height=3.5cm]{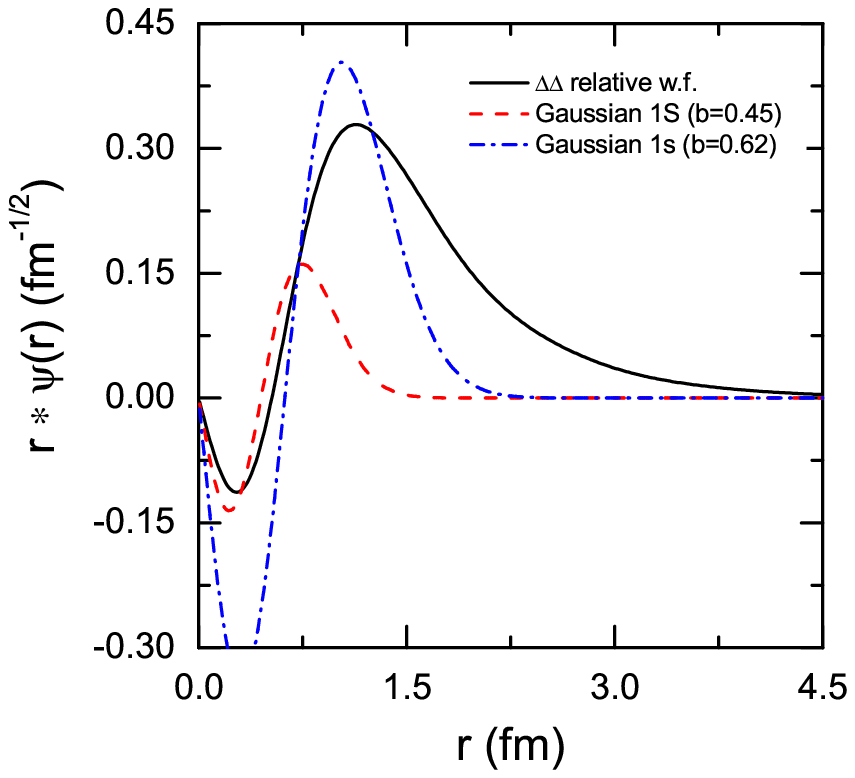}
\caption{(Color online) Wave functions in the TCC calculation. The
left panel shows the $^7S_3$ wave function of the CC channel, and
the right panel presents the remained wave function after
subtracting a $\psi_1$ wave function and a Gaussian function with
(1s) excitation.} \label{wf}
\end{figure}

{\label{WF}}
Now, we select the trial wave function of the six-quark system in
the model space of SCC. Since the ground state of the six-quark
system with $(IS)=(03)$ and $L=0$ in this model space has the
symbolic form of
\begin{eqnarray}
\psi_1=\left((0s)^6[6]_{orb}[111111]_{SFC}\right)_{(IS)=(03)},
\end{eqnarray}
where $(0s)$ represents a $(0s)$ orbital wave function of the
harmonic oscillator with $b_1$ being the size parameter, the total
orbital wave function has a [6] symmetry, and the total wave
function in the spin-flavor-color space has a [111111] symmetry. As
shown in Ref.~\cite{Liumsh:2015}, this configuration is not adequate
to describe this system, therefore the components with higher
excitations should be included. On the other hand, the result in the
TCC calculation shows that the CC component in the wave function of
$d^*$ has a rather large fraction, about 2/3. By re-organizing such
a wave function (refer to the right panel of Fig.~\ref{wf} in
Ref.\cite{Huang:2014kja}) to form a
$((0s)^6[6]_{orb}[111111]_{SIC})_{(IS)=(03)}$ type wave function,
one sees that the obtained wave function has a large fraction of
about $80$\% in the total wave function. It implies that the main
character of the observed structure by the WASA-at-COSY
collaboration should also appear in the SCC calculation. If the
structure proposed in our previous TCC calculation is reasonable,
the curve in the right panel of Fig.~\ref{wf} tells us that one
needs at least an additional component of the wave function with one
node in radial, namely a (1s) radially excited wave function.
Thus, in the lowest order approximation, we could adopt an
additional wave function which has $2\hbar \omega$ excitation to
supplement the inadequacy of $\psi_1$ in describing $d^*$. Now, we
pick up all the wave functions which have $2\hbar\omega$ radial
excitation
\begin{eqnarray*}
\left((0s)^5(1s)[6]_{orb}[111111]_{SIC}\right)_{(IS)=(03)},
\end{eqnarray*}
and
\begin{eqnarray*}
\left((0s)^4(0p)^2[6]_{orb}[111111]_{SIC}\right)_{(IS)=(03)},
\end{eqnarray*}
where $(0p)$ and $(1s)$ denote the orbital wave functions of a quark
moving in the $(0p)$ and $(1s)$ orbits, respectively, which also
take the harmonic oscillator form with a size parameter of $b_2$.
Their orbital parts can be written explicitly as
\begin{eqnarray*}
(0s)^5(1s)[6]_{orb} = \sqrt{\frac{1}{6}}\sum_{i=1}^{6}[(0s)^5(1s)_i],
\end{eqnarray*}
and
\begin{eqnarray*}
(0s)^4(0p)^2[6]_{orb} = \sqrt{\frac{1}{15}}\sum_{i<j}^{6}[(0s)^4(0p)^2_{ij}].
\end{eqnarray*}
Their linear combination could form a configuration which orthogonal
to $\psi_1$ and the excited wave function of the center of mass
motion (CM), as long as $b_1=b_2$. Then, the supplemented
configuration can be taken as
\begin{eqnarray}
\psi_2 &=& \left[\left(  \sqrt{\frac{5}{6}} (0s)^5(1s)[6]_{orb}\right.\right. \nonumber \\
&+& \left.\left. \sqrt{\frac{1}{6}} (0s)^4(0p)^2[6]_{orb}\right)
[111111]_{SIC}\right]_{(IS)=(03)},\nonumber \\
\end{eqnarray}
where the size parameter $b_2$ is chosen as a variational parameter
(later, $b_1$ can be varied as well for getting an even more stable
solution).

Finally, the trial wave function can be expressed as
\begin{eqnarray}
\Psi_{6q} &=& c_1  \psi_1 +c_2 \psi_2,
\end{eqnarray}
where $c_1$ and $c_2$ are the mixing coefficients. It should
particularly be emphasized that the $\psi_1$ and $\psi_2$ are not
orthogonal in general, except $b_2=b_1$.

\section{Results and Discussions}{\label{results}}

\begin{figure}[htbp]
\centering
\includegraphics [width=8.0cm, height=5.5cm]{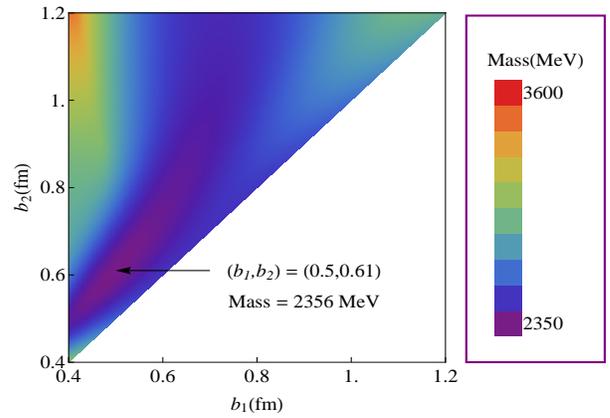}
\caption{(Color online) The $b_1$ and $b_2$ dependence of the
$d^*(2380)$ mass.} \label{mass1}
\end{figure}

As mentioned in the previous section, because $\psi_2$ is employed
to complement $\psi_1$, $\psi_2$ with a $2\hbar\omega$ excitation
can be regarded as an inner structural deformation of the concerned
six-quark system. Generally, the size of $\psi_2$ should be larger
than $\psi_1$'s, namely $b_2>b_1$, is required. The eigenvalue
problem of the SCC of the six-quark system with given values of
$b_1$ and $b_2$ should be considered first. Due to non-orthogonality
of $\psi_1$ and $\psi_2$, a generalized eigenvalue equation, secular
equation,
\begin{eqnarray}
\sum_{j=1}^2\,\langle \,\psi_{i} \mid H \mid \psi_{j}\,
\rangle\,c_j~=~E~\sum_{j=1}^2\,\langle \,\psi_{i} \mid \psi_{j}\,
\rangle\,c_j\nonumber\\
~~~~~~~~~~~(i=1,2)
\end{eqnarray}
should be solved with certain values of $b_1$ and $b_2$, for
instance $b_1=0.5fm$, as in the TCC calculation, and a value greater
than $0.5fm$ for $b_2$. Changing value of $b_2$, the obtained
eigenvalue becomes a function of $b_2$. The actual value of $b_2$
should be achieved by the variational procedure, so that the system
would have minimum mass.  To ensure the system being even more
stable, $b_1$ should further be regarded as a changeable parameter,
namely a two-parameter variation
\begin{eqnarray}
\frac{\partial^2 \,\langle \,\Psi_{6q} \mid H \mid \Psi_{6q}\,
\rangle}{\partial\,b_1~\partial\,b_2}~=~0, \\ \nonumber
\end{eqnarray}
should be performed. Now, the obtained mass of $d^*$ depends on the
size parameters $b_1$ and $b_2$. We plot such a dependence in
Fig.~\ref{mass1}.

From this figure, one sees that there do exist a stable point with
respect to $b_1$ and $b_2$ in the $b_1$-$b_2$ plane, where
$b_1=0.5~\rm{fm}$, $b_2=0.61~\rm{fm}$, and the mass of the system
reaches its minimum value of $2356 ~\rm{MeV}$. One also finds that
although we variate $b_1$ and $b_2$ simultaneously, the outcome
value for $b_1$ is very close to its starting value which has been
used in the determination of model parameters. It reflects the
reliability of the result.

For comparison, we also calculate the mass of the system with
$b_1=0.5fm$ when the chiral field induced interaction $V^{ch}_{ij}$
is absent. The obtained mass is about $2481$MeV. A comparison
between these two masses shows an implication that by taking into
account the potentials induced by the chiral fields, the effect of
NPQCD in the medium-range should be reasonably included. As a
consequence, the single cluster six-quark system with the orbital
[6] symmetry is indeed bound with respect to the threshold of
$\Delta\Delta$, and the mass of system is close to the experimental
data, which is contradict the conclusion in Ref.~\cite{Park:2015nha}
where the important medium-range interaction due to the chiral
symmetry consideration is missing. This also means that such a
system is likely a hexaquark dominated state.

Moreover, we obtain the wave function of the bound state with its
coefficients $c_1$ and $c_2$ being 0.849 and -0.727, respectively.
Using this wave function, we in principle can estimate the decay
width of the obtained state $d^*$. However, the obtained width is
too small to explain the data. This is because that the hypothetical
trial wave function $\Psi_{6q}$ could not well approach to the
reality of the observed structure due to an improper cut in adopting
inner structural deformation wave functions. Actually, we find that
the wave functions with a size parameter of $0.5-1.1fm$ are the most
important pieces which would provide the major contribution to the
width. However, in order to use a simplest possible model to
describe the binding behavior of the system without losing major
character, the pieces describing the information of the inner
structural deformation with larger size parameters are absent in our
hypothetical trial wave function, namely the adopted harmonic
oscillator form of the single cluster trial wave function cannot
properly describe the real behavior of the system at surface region.
A sophisticated study should be carried out further. Nevertheless,
our model with a rather simple but meaningful six quark structure, a
basic ground state component with the spatial [6] symmetry plus an
inner structural deformation component with a $2\hbar\omega$
excitation which is also in the spatial [6] symmetry, still gives
the main characters of $d^*$, in this case, its mass is about
$46$MeV higher than the $\Delta N\pi$ threshold but about $108$MeV
lower than the $\Delta\Delta$ threshold, although its mass is about
$24$MeV smaller than the observed value, and its width is much
smaller than the width of two $\Delta$'s width of about $230$MeV.

It should be noted that the $2\hbar \omega$ excited state may have
another symmetry structure, i.e., a wave function with a spatial
[42] symmetry. However, because the wave functions with spatial [6]
and [42] symmetries are orthogonal to each other, and the central
force reserves the spatial symmetry [6], inclusion of the [42]
symmetry wave function will not affect final result. Therefore, we
disregard such a configuration in this preliminary calculation. We
also limit our discussion on the states with excited energy higher
than $2\hbar \omega$, due to their larger kinetic energies, and
consequently, less influence on the mass of $d^*(2380)$.

\begin{figure}[htbp]
\centering
\includegraphics [width=8.0cm, height=5.5cm]{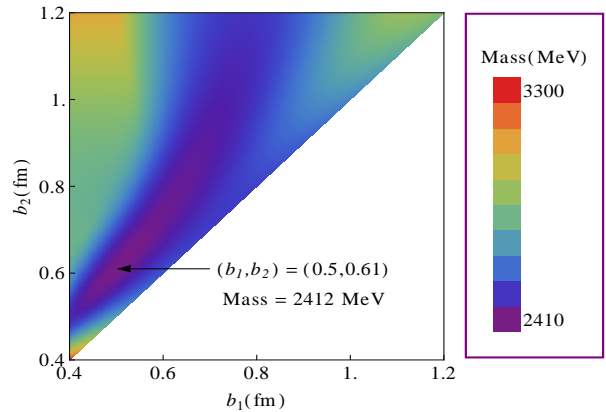}
\caption{(Color online) The $b_1$ and $b_2$ dependence of the
$IS=30$ state.} \label{mass2}
\end{figure}

The mirror state of $d^*$ whose quantum numbers are $IS=30$ is
calculated with the same trial wave function as well. We plot the
mass dependence on $b_1$ and $b_2$ in Fig.~\ref{mass2}. The stable
point occurs at $b_1 = 0.5~fm$ and $b_2=0.61~fm$ with a mass of
$2412$MeV, which is around the $\Delta \Delta$ threshold.

\section{Summary}{\label{summary}}

In this paper, the intrinsic binding behavior of the experimentally
observed $d^*(2380)$ are studied in the SCC approximation within the
chiral SU(3) constituent quark model. For simplicity but keeping
major character, the trial wave function is chosen as a combination
of a basic ground state component with the spatial [6] symmetry and
an inner structural deformation component with a $2\hbar\omega$
excitation which is also in the spatial [6] symmetry. A
two-parameter variational calculation is performed in order to reach
a stable state where the mass of the system has a minimum value. It
is shown that there do exist a stable point, the corresponding $b_1$
and $b_2$ are $0.5fm$ and $0.61fm$, respectively, and the energy is
about $2356MeV$, which is qualitatively consistent with the observed
value. This result contradicts that in Ref.~\cite{Park:2015nha}
where the interaction between quarks involves OGE and confinement
potentials only. This is because that in the QCD inspired
constituent quark model, the mass of the constituent quark comes
from the restoration of the the chiral symmetry, thus the chiral
symmetry must be considered. As a practical way, the chiral field
induced potential which describes the medium-range NPQCD effect
should be introduced into the constitute quark model. That is why
with our chiral SU(3) constitute quark model, a much lower mass of
the six-quark system can be obtained. Moreover, although the
estimated decay width of the system does not contradict the data, it
is too small to match the observed value, because the hypothetic
trial wave function is too simple to describe inner structural
deformation of the system especially in the surface region where the
contribution from the tail of the wave function of the system
dominates the width. In a word, the binding behavior of a single
cluster six-quark system with $(IS)=(03)$ is compatible with the
result from the RGM calculation qualitatively. This means that the
hexaquark dominated picture may be a promising picture for $d^*$.
For completion, the mirror state of $d^*$ is also studied. The mass
of this state is about $2412MeV$ which is around the $\Delta\Delta$
threshold.

\bigskip
\noindent
\begin{center}
{\bf ACKNOWLEDGEMENTS}\\
\end{center}

We would like to thank Qiang Zhao and H. Clement for their useful
and constructive discussions. This project is partly supported by
the National Natural Science Foundation of China under Grants
Nos.~10975146, 11165005, 11475181, 11475192, 11565007 and 11521505, and the
IHEP Innovation Fund under the No. Y4545190Y2, as well as
supported by the DFG and the NSFC through funds (11621131001) provided to the Sino-Germen CRC 110 ¡°Symmetries
and the Emergence of Structure in QCD¡±. Support by the China Postdoctoral Science Foundation under Grant No. 2016M601133
is also appreciated.

\end{document}